\documentstyle[prl,aps,twocolumn,epsf]{revtex}

\newcommand{\beq}{\begin{equation}}
\newcommand{\eeq}{\end{equation}}
\newcommand{\bea}{\begin{eqnarray}}
\newcommand{\eea}{\end{eqnarray}}
\newcommand{\eps}{\epsilon}
\newcommand{\veps}{\varepsilon}

\newcommand{\nn}{\nonumber}
\newcommand{\benn}{\begin{displaymath}}
\newcommand{\eenn}{\end{displaymath}}
\newcommand{\ket}[1]{| #1 \rangle}                     %  | >
\newcommand{\bra}[1]{\langle #1 \, |}                  %  < |
\begin{document}

\title{\bf \LARGE Local Density Approximation for Systems with Pairing
Correlations }
\vspace{0.50cm}
\author{Aurel Bulgac }
\vspace{0.50cm}
\address{Department of Physics, University of
Washington, Seattle, WA 98195--1560, USA}
\maketitle

% \vspace{10mm}

\begin{abstract}

I formulate a local density approximation for fermion systems with
pairing correlations based on a rapidly converging renormalization
scheme for the pairing field.

\end{abstract}

\draft

% 21.60.Jz Hartree-Fock and random phase approximations
% 21.60.-n Nuclear-structure models and methods
% 21.30.Fe Forces in hadronic systems and effective interactions

% 71.15.Mb Density functional theory, local density approximation,
%       gradient and other corrections
% 74.20.Fg BCS theory and its development

% \vspace{10mm}

%\today

The proof that for any fermion system there exist a unique energy
density functional of the density matter distribution $\rho(\bbox{r})$
alone, namely $E_{gs}=\mathrm{min} \; \int d^3r
{\cal{E}}_{DFT}(\rho(\bbox{r}))$, is disarmingly simple \cite{hk}.
However, except for some trivial cases, the exact form of this
functional is still a mystery and no constructive algorithms for its
determination have been suggested so far. Significant progress has
been achieved however within the Kohn--Sham Local Density
Approximation (LDA) of the Density Functional Theory (DFT). For a
normal fermion system (with no pairing correlations) Kohn and Sham
have shown that the ground state energy of any fermion system is a
functional of its kinetic energy and matter density distributions,
namely $E_{gs}=\mathrm{min} \; \int d^3r {\cal{E}}_{LDA}
(\tau(\bbox{r}), \rho(\bbox{r}) )$.  The current philosophy is that
one should determine this functional from homogeneous infinite matter
calculations and then use it to describe properties of either infinite
inhomogeneous or finite systems \cite{hk}. By the same token, one
would expect that the formulation of a LDA for fermion systems with
pairing correlations should be straightforward and that a
corresponding universal LDA energy density functional ${\cal{E}}_{LDA}
(\tau ,\rho ,\nu )$ of the kinetic energy $\tau(\bbox{r})=2\sum _i
|\bbox{\nabla} v_i(\bbox{r})|^2$, normal $\rho(\bbox{r})=2\sum _i
|v_i(\bbox{r})|^2$ and anomalous $\nu (\bbox{r})=\sum _i
v_i^*(\bbox{r})u_i(\bbox{r})$ densities exists.  (I shall be concerned
here explicitly with the case of $s$--pairing in the so called weak
coupling limit. Generalizations seem possible however.)  The LDA
extension described in Refs. \cite{oliveira} is in terms of the
anomalous density matrix $\nu
(\bbox{r}_1,\bbox{r}_2)=\bra{gs}\hat{\psi}_\uparrow
(\bbox{r}_1)\hat{\psi}_\downarrow (\bbox{r}_2)\ket{gs}$.  Upon
variation of the quasi--particle wave functions $v_i(\bbox{r}),
u_i(\bbox{r})$ under standard restrictions one obtains the Kohn--Sham
equations, with a structure identical to the Hartree--Fock--Bogoliubov
(HFB) or Bogoliubov--de Genes (BdG) equations:
%====================================================================
\bea
& &  [h (\bbox{r})  - \mu] u_E (\bbox{r})
     + \Delta (\bbox{r})  v_E (\bbox{r})
    = E u_E (\bbox{r}) , \label{eq:hfb0u}\\
& &  \Delta^* (\bbox{r}) u_E (\bbox{r})  -
    [ h^* (\bbox{r}) - \mu ] v_E (\bbox{r})
     = E v_E (\bbox{r}),  \label{eq:hfb0v}
\eea
%====================================================================
where $ h (\bbox{r}) $ is the single--particle hamiltonian, $\Delta
(\bbox{r})= -\frac{\delta E_{gs}}{\delta \nu^*(\bbox{r})}$ and $\mu$
is the chemical potential. Each quasi--particle state could be
characterized by additional quantum numbers besides the
quasi--particle energy $E$, which I shall not explicitly display
however. In all the formulas presented here I shall likewise not
display the spin degrees of freedom. One can show that the mere
locality of the pairing field $\Delta (\bbox{r})$ leads to a divergent
diagonal part of the anomalous density $\nu (\bbox{r},\bbox{r})$
\cite{ab,bruun,abyy}. When $|\bbox{r}_1-\bbox{r}_2|\rightarrow 0$ the
anomalous density matrix has the singular behavior $\nu
(\bbox{r}_1,\bbox{r}_2)=\sum _{E\ge 0}
v_E^*(\bbox{r}_2)u_E(\bbox{r}_1) \propto \frac{
1}{|\bbox{r}_1-\bbox{r}_2|} $.  As a result, the local
self--consistent pairing field $\Delta (\bbox{r})$ cannot be defined.
(When summing over the spectrum, the sum becomes an integral if the
spectrum is continuous and vice versa for an integral. I shall be
casual in using either a summation or integration notation, hoping
that the context makes this distinction obvious.)  The existence of
this particular divergence was the main obstacle in introducing an
extension of the LDA approach to systems with pairing correlations.
Fortunately, this divergence is one more example of the infinities
which infest Quantum Field Theory (QFT) and for which the techniques
to regularize them in a controlled fashion exist and can be extended
and applied to inhomogeneous systems as well now.

It is instructive to show how this divergence emerges and the simplest
system to illustrate this is an infinite homogeneous one. Since the
divergence is due to high momenta, thus small distances
$|\bbox{r}_1-\bbox{r}_2|$, this type of divergence is universal and
has the same character in both finite and infinite systems.  Until
recently methods to deal with this divergence were known only for
infinite homogeneous systems
\cite{yang,gorkov,randeria,mohit,fayans,george,hsu} and only recently
ideas were put forward on how to implement a renormalization scheme
for the case of finite or inhomogeneous systems
\cite{bruun,abyy}. Assuming for the sake of simplicity that the
spectrum of the HF operator is simply $\veps
(\bbox{k})=\hbar^2k^2/2m$, one can represent the anomalous density
matrix as follows \cite{ab,bruun,abyy}
%------------------------------------------------------------------
\bea
& & \nu(\bbox{r}_1,\bbox{r}_2)=
\int \frac{d^3k}{(2\pi)^3} 
\frac{\exp[i\bbox{k}\cdot(\bbox{r}_1-\bbox{r}_2)]\Delta }{
2\sqrt{[\veps (\bbox{k}) -\mu]^2+\Delta ^2}} \\
& & \equiv
\int \frac{d^3k}{(2\pi)^3} \exp[i\bbox{k}\cdot(\bbox{r}_1-\bbox{r}_2)]
\left \{
\frac{\Delta }{2\sqrt{[\veps (\bbox{k}) -\mu]^2+\Delta ^2}} \right .\nn \\
& &\left . -\frac{\Delta}{2[\veps (\bbox{k})-\mu  -i\gamma ]}\right \} 
+ \frac{\Delta m \exp ( ik_F|\bbox{r}_1-\bbox{r}_2|) }{
4\pi \hbar ^2|\bbox{r}_1-\bbox{r}_2| },  \label{eq:nu}
\eea
%------------------------------------------------------------------
where $\mu = \hbar ^2k_F^2/2m  $.  The last integral expression is
well defined for all values of the coordinates $\bbox{r}_{1,2}$. Once
one has recognized the existence of a divergence, the next step is to
devise a way to regularize the theory. In a nutshell, what one has to
do is to subtract the divergent part 
$ \frac{\Delta m }{4\pi \hbar ^2|\bbox{r}_1-\bbox{r}_2| } $
from the rest in the limit
$|\bbox{r}_1-\bbox{r}_2|\rightarrow 0$. Formally one can justify
this apparently rather arbitrary procedure, either by following the
steps outlined typically in renormalizing the gap equation in infinite
systems -- by relating the divergent part with the scattering amplitude
\cite{yang,gorkov,randeria,mohit,fayans}-- or by using
well--known approaches in QFT -- for example
dimensional regularization \cite{george,hsu}, or another QFT approach
of introducing appropriate counterterms with explicit cut--offs -- or
one can follow the philosophy of the pseudopotential approach
\cite{bruun,abyy,blatt}. In all cases one naturally arrives at the
same final value for the gap. The renormalized gap equation can be
written as
%------------------------------------------------------------------
\bea
  & & -\frac{1}{g} =  \int \frac{d^3k}{(2\pi)^3}
\left [ 
  \frac{1 }{2\sqrt{[\veps (\bbox{k}) -\mu]^2+\Delta ^2}} 
\right . \nn \\
& & \left .  
-\frac{1}{2[ \veps (\bbox{k}) -\mu - i\gamma ]}
\right ]
  +\frac{ik_Fm }{4\pi \hbar^2},  \label{eq:gapinf}
 \eea
%------------------------------------------------------------------
where the coupling constant $g$ is defined as
%------------------------------------------------------------------
\beq
g\delta(\bbox{r}_1-\bbox{r}_2)=
\frac{\delta ^2 E_{gs} }{\delta \nu^*(\bbox{r}_1)\delta \nu(\bbox{r}_2) }.
\eeq
%------------------------------------------------------------------
Previous approaches
\cite{yang,gorkov,randeria,mohit,fayans,george,hsu} use
$\veps(\bbox{k})$ only in the second term under the integral and in
that case the last imaginary term does not appear.  I have assumed
here the simplest dependence of the LDA energy density functional on
the anomalous density $\nu(\bbox{r})$, namely ${\cal{E}}(\tau
(\bbox{r}) , \rho (\bbox{r}),|\nu(\bbox{r})|^2)$, merely for the sake
of the simplicity of the presentation, but more general forms can be
used as well. A note of caution: it would be incorrect to interpret
some of the above formulas in the same manner as similar looking
formulas appearing in various treatments of the pairing correlations
with a zero--range interaction $V(\bbox{r}_1-\bbox{r}_2)=
g\delta(\bbox{r}_1-\bbox{r}_2)$ (which can be related with the zero
energy two--particle scattering amplitude $g=4\pi\hbar^2a/m$). As it
is well known for quite some time, even in the low density region,
when $k_F|a|\ll 1$, there are significant medium polarization
corrections to the pairing gap \cite{gor}. The present LDA treatment
is not limited by similar restrictions on the density. In the LDA
energy density functional the polarization effects are already
implicitly included in the definition of ${\cal{E}}(\tau(\bbox{r}),
\rho(\bbox{r}),|\nu(\bbox{r})|^2)$ and the coupling constant $g$ has
no simple and direct relation to the vacuum two--particle scattering
amplitude $a$.  In this sense the LDA is similar in spirit to the
Landau fermi liquid theory.

Eq. (\ref{eq:gapinf}) can be used to extract from known properties of
homogeneous infinite matter (such as $\veps (\bbox{k})$, $\Delta$ and
density) the specific value of the coupling constant $g$ to be used in
constructing
${\cal{E}}(\tau(\bbox{r}),\rho(\bbox{r}),|\nu(\bbox{r})|^2)$.
Assuming that a full microscopic calculation of homogeneous matter at
a given density $\rho= k_F^3/3\pi^2$ has been performed and that the
value of the pairing gap at the fermi level is known, one can, using
Eq. (\ref{eq:gapinf}), calculate directly $g(\rho )$ and thus obtain
the simplest approximation to the LDA energy density functional
${\cal{E}}_{LDA}(\tau(\bbox{r}),\rho(\bbox{r}),|\nu(\bbox{r})|^2) =
{\cal{E}}_0(\tau(\bbox{r}),\rho(\bbox{r}))+ g(\rho(\bbox{r})) |
\nu(\bbox{r}) |^2$, where ${\cal{E}}_0(\tau(\bbox{r}),\rho(\bbox{r}))$
is the Kohn--Sham energy density functional in the absence of pairing
correlations.  In many treatments of the pairing correlations in
infinite systems authors often underline the dependence of the pairing
gap on momentum, that is $\Delta (\bbox{k})$. On one hand, typical
calculations \cite{elgaroy} of the pairing field $\Delta (\bbox{k})$
in infinite systems (with no medium polarization effects taken into
account so far) show that for large momenta the pairing field
decreases, as one would naturally expect.  On the other hand, as soon
as the momentum of a quasiparticle state is sufficiently different
from the fermi momentum, when $|k-k_F| \approx m \Delta (k_F)/\hbar^2
k_F\ll k_F$,
 the effect of the pairing correlations on the single--particle
properties is small, if not negligible. To a very good accuracy
$E(\bbox{k})=\sqrt{(\veps(\bbox{k})-\mu)^2+\Delta^2(\bbox{k})} \approx
\sqrt{(\veps(\bbox{k})-\mu)^2+\Delta^2(\bbox{k}_F)}\approx
|\veps(\bbox{k})-\mu|$ and thus the use of a $k$--independent pairing
field is a fair approximation. This is just another way of stating
that the size of the Cooper pair $\hbar^2 k_F/m\Delta$ \cite{mohit} is
much larger then the average interparticle separation in the weak
coupling limit. Typically this takes place when also the range of the
pairing interaction is smaller than the size of the Cooper pair as
well, and thus the pairing interaction could be described by a single
coupling constant.

Even though apparently the divergence has been successfully dealt with
(in infinite homogeneous systems), a closer inspection of the entire
approach reveals an inconsistency, which is somewhat hard to spot. The
divergence is due to high momenta and for that reason one has
subtracted the term $\Delta /2[\veps(\bbox{k})-\mu -i\gamma ]$ in Eqs.
(\ref{eq:nu},\ref{eq:gapinf}). Far away from the fermi surface
however, the problematic term $\Delta/2\sqrt{[\veps (\bbox{k})
-\mu]^2+\Delta ^2}$ behaves rather like $\Delta
/2|\veps(\bbox{k})-\mu|$ instead. The main difference between these
two subtraction procedures appears for hole--like states.  As the
fermi energy is finite, the integral over states below the fermi level
is also finite.  This feature, which breaks the approximate symmetry
between the particle and hole states, is rather unsatisfactory and it
has no theoretical underpinning. On one hand, in calculating the
integral over the single--particle spectrum above the fermi level one
expects a relatively fast convergence, when the energy of the particle
states is a ``few gaps $\Delta$ away''. On the other hand, the
integral over the hole states converges only for energies of the order
of the fermi energy $\eps_F=\hbar ^2k_F^2/2m$. Clearly, in most cases
of interest, the so called weak coupling limit, when $\Delta \ll
\eps_F$, there is absolutely no physical reason to take into account
single--particle states so far away from the fermi level in order to
describe global or meanfield properties of nuclei in particular.

\begin{figure*}[h,t,b]

\begin{center}
\epsfxsize=6.0cm
\centerline{\epsffile{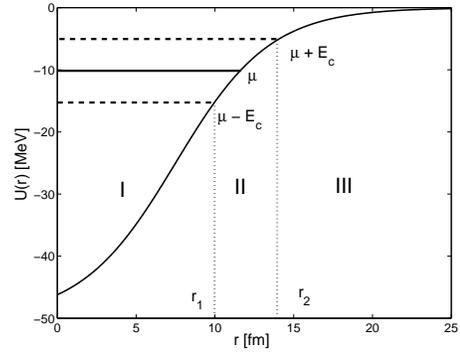}}
\end{center}

\caption{ In region I all three wave vectors
$l_c(\bbox{r}),k_F(\bbox{r})$ and $k_c(\bbox{r})$ are real and both
subtraction terms are present. In region II $l_c(\bbox{r})$ is
imaginary and the corresponding subtraction term in
Eq. (\ref{eq:nurenorm}) should be dropped.  In region III all three
wave vectors are imaginary and both subtraction terms in
Eq. (\ref{eq:nurenorm}) should be dropped.  Even though in region II
$k_F(\bbox{r})$ becomes imaginary for larger $r$, the corresponding
subtraction term containing $k_F(\bbox{r})$ is real everywhere and it
should be retained. }

\label{fig:fig1}

\end{figure*}

I show here how a relatively simple regularization scheme can easily
deal with this problem in a very clear and easily implementable
manner, suitable for any system, finite or infinite, homogeneous or
inhomogeneous.  The regularized anomalous density is calculated from
the following expression:
%--------------------------------------------------------------
\bea
& & \nu_{reg} (\bbox{r}) :=
 \int _0^{E_c} dE g_{H\!F\!B}(E)  v^*_E(\bbox{r})u_E (\bbox{r})  \\
& &- \int^{\mu+E_c}_{\mu-E_c} d\veps g_{H\!F}(\veps)\frac{\Delta(\bbox{r})}{2}
  \frac{\psi^*_\veps(\bbox{r})\psi_\veps(\bbox{r})}{|\veps-\mu|+i\gamma} 
  +\frac{\Delta(\bbox{r})}{2} \Gamma  ^{reg} (\bbox{r},\mu) , \nn
\eea
%---------------------------------------------------------------
where $g_{H\!F\!B}(E)$ and $g_{H\!F}(\veps)$ are the HFB and HF density of
states respectively,
%====================================================================
\bea
& & [h(\bbox{r})-\veps ]\psi_\veps(\bbox{r})=0, \\
& & \Gamma (\bbox{r}_1,\bbox{r}_2,\mu)=
\int d\veps g_{H\!F}(\veps)
\frac{\psi^*_\veps(\bbox{r}_1)\psi_\veps(\bbox{r}_2)}{|\veps-\mu|+i\gamma}\\
& & = \frac{m}{ 2\pi \hbar ^2|\bbox{r}_1-\bbox{r}_2| } +
\Gamma ^{reg}(\bbox{r},\mu)  +
{\cal{O}}(|\bbox{r}_1-\bbox{r}_2|).
\eea
%====================================================================
$\gamma$ is as usual a small infinitesimal quantity and
$\bbox{r}=\bbox{r}_{1,2}$ in the limit $|\bbox{r}_1-\bbox{r}_2|\rightarrow 0$. 
As in Ref. \cite{abyy}, I
shall use a Thomas--Fermi approximation for the single--particle wave
functions $\psi_\veps(\bbox{r})$ and energies in order to evaluate the
regulator. After introducing the
local wave vectors $l_c(\bbox{r}) \le k_F(\bbox{r}) \le k_c(\bbox{r})$
%====================================================================
\bea
& & h(\bbox{r})=-\frac{\hbar^2\bbox{\nabla}^2}{2m}+U(\bbox{r}),\\
& &
\left [ \frac{\hbar^2k_c^2(\bbox{r})}{2m}+ U(\bbox{r})\right ]=E_c +\mu ,\\
& &
\left [ \frac{\hbar^2l_c^2(\bbox{r})}{2m}+ U(\bbox{r})\right ]=-E_c+\mu ,\\
& & \frac{\hbar^2k_F^2(\bbox{r})}{2m}+ U(\bbox{r})=\mu
\eea
%====================================================================
and after some straightforward manipulations one can show that the
renormalized anomalous density introduced above acquires the following
form
%====================================================================
\bea
& & \nu_{reg}(\bbox{r}) :=
\int_0^{E_c} dE g_{H\!F\!B}(E)  v^*_E(\bbox{r})u_E (\bbox{r}) 
  \label{eq:nurenorm}\nn \\
& & -\frac{\Delta(\bbox{r})m k_c(\bbox{r})}{2\pi^2\hbar ^2}
\left \{ 1
  -\frac{k_F(\bbox{r})}{2 k_c(\bbox{r})}
\ln \frac{k_c(\bbox{r})+k_F(\bbox{r})}{k_c(\bbox{r})-k_F(\bbox{r}) }
    \right \}  \nn \\
& &     -\frac{\Delta(\bbox{r})m l_c(\bbox{r})}{2\pi^2\hbar ^2}
\left \{ 1
    -\frac{k_F(\bbox{r})}{2 l_c(\bbox{r})}
\ln \frac{k_F(\bbox{r})+l_c(\bbox{r})}{k_F(\bbox{r})-l_c(\bbox{r}) }
     \right \}    .  \label{eq:nureg}
\eea
%====================================================================
The only formal difference between this expression and the
corresponding expression introduced in Ref. \cite{abyy} is in the
terms containing the second cut--off momentum $l_c(\bbox{r})$ (last line).  
If either one of the wave vectors $l_c(\bbox{r})$ or $k_c(\bbox{r})$
becomes imaginary, then the corresponding terms in the renormalized
anomalous density $\nu _{reg}(\bbox{r})$ should be dropped. However, if
the wave vector $k_F(\bbox{r})$ becomes imaginary, the renormalized
anomalous density is real and the above definition should be used, see
Fig. \ref{fig:fig1} for a generic situation.

It is convenient to introduce a notation for the cut--off anomalous density
$\nu_c (\bbox{r}) :=
 \int _0^{E_c} dE g_{H\!F\!B}(E)  v^*_E(\bbox{r})u_E (\bbox{r})$
and an effective position running coupling constant 
%--------------------------------------------------------------
\bea
& & \frac{1}{ g_{\mathit{eff}}(\bbox{r})}=
\frac{1}{g}
 -\frac{m k_c(\bbox{r})}{2\pi^2\hbar ^2}
\left \{ 1
  -\frac{k_F(\bbox{r})}{2 k_c(\bbox{r})}
\ln \frac{k_c(\bbox{r})+k_F(\bbox{r})}{k_c(\bbox{r})-k_F(\bbox{r}) }
    \right \} \nn \\
& &     -\frac{m l_c(\bbox{r})}{2\pi^2\hbar ^2}
\left \{  1
    -\frac{k_F(\bbox{r})}{2 l_c(\bbox{r})}
\ln \frac{k_F(\bbox{r})+l_c(\bbox{r})}{k_F(\bbox{r})-l_c(\bbox{r}) }
     \right \}. \label{eq:geff}
\eea
%---------------------------------------------------------------
In the limit $k_F \rightarrow 0$  (in vacuum) the value of this effective 
running coupling constant agrees with that derived in Ref. \cite{henning}.
Using these notations one obtains for the renormalized pairing field
%----------------------------------------------------------------------
\beq
\Delta (\bbox{r}) = -g\nu_{reg}(\bbox{r})=
   -g_{\mathit{eff}}(\bbox{r}) \nu_c (\bbox{r}) .
\label{eq:Delta}
\eeq
%----------------------------------------------------------------------

%--------------------------------------------------------------------
%--------------------------------------------------------------------

\begin{figure*}[h,t,b]

\begin{center}
\epsfxsize=6.0cm
\centerline{\epsffile{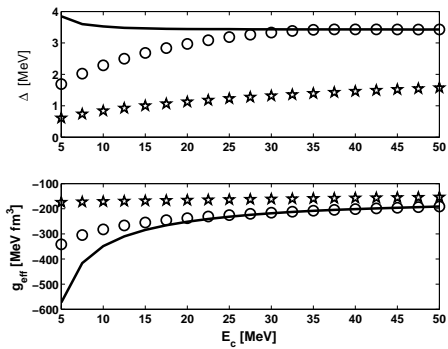} } 
\end{center}
\caption{ The gap $\Delta$  and the effective coupling
  constant $g_{\mathit{eff}}$ as a function of the
  cut-off energy $E_c$ for three regularization schemes. The full lines
  correspond to  
  calculations using  Eqs. (\ref{eq:nureg} -- \ref{eq:Delta}). 
  Circles correspond to the regularization scheme presented in 
  Ref.  [5] (when only terms with $k_c$ are present). 
  The pentagrams correspond to the vacuum regularization scheme [16].
  The calculation was performed for
  homogeneous neutron matter with $\rho = 0.08\;{\mathrm{fm}}^{-3}$ and 
  $g=-250 \;{\mathrm{MeV}}\cdot{\mathrm{fm}}^3$. }

\label{fig:fig2}

\end{figure*}

Even though the cut--off momenta $k_c(\bbox{r})$ and $l_c(\bbox{r})$
and the cut--off quasiparticle energy $E_c$ explicitly appear in the
definition of both the effective coupling constant and of the cut--off
anomalous density, the gap $\Delta (\bbox{r})$ is indeed cut--off
independent, once the cut--off energy $E_c$ has been taken
sufficiently far from the fermi surface. This situation is similar to
the situation described in Ref. \cite{abyy}, with the single
difference that in the present case the convergence is achieved for
significantly smaller values of $E_c$.  As one can judge from
Fig. \ref{fig:fig2}, the present regularization scheme is indeed very
fast converging, while the regularization scheme presented in
Ref. \cite{abyy} converges as expected at energies of the order of the
fermi energy $\epsilon _F=\hbar^2k_F^2/2m$.  At the same time, the
traditional approach based on a $\delta$--function with cut-off energy
$E_c$ \cite{henning} (for which $g_{ \mathit{eff}} (\bbox{r})
=g/[1-gmk_c(\bbox{r})/2\pi^2\hbar^2]$) converges extremely slowly, and
even at $E_c=1000$ MeV is still about 20\% off the converged value.

When computing the total energy of such a system one has to be careful
and evaluate
%--------------------------------------------------------------------
\bea
& & E_{gs} = \int d^3r \left [ \frac{\hbar^2}{2m}\tau_c(\bbox{r}) -
\Delta (\bbox{r})\nu _c(\bbox{r})\right ] \nn \\ 
& & + \int d^3r {\cal{E}}_0(0,\rho (\bbox{r})) , 
\eea
%--------------------------------------------------------------------
where the kinetic energy density is evaluated as $\tau_c (\bbox{r})
:=2 \int _0^{E_c} dE g_{H\!F\!B}(E)|\bbox{\nabla} v_E(\bbox{r})|^2$.
(I have assumed here the simplest dependence of $
{\cal{E}}_0(\tau,\rho )$ on $\tau$.)  Only this combined expression,
containing the trace of the kinetic energy with the trace of the
pairing field and of the cut--off anomalous density, is converging as
a function of the cut--off quasiparticle energy $E_c$
\cite{george}. The reason is that $\tau_c (\bbox{r})$ diverges in a
similar manner as $ \nu_c (\bbox{r})$ as a function of $E_c$.

%--------------------------------------------------------------------
%--------------------------------------------------------------------

The formalism described here paves the way to a LDA to pairing in the
spirit of the Kohn--Sham theory \cite{hk} . One has simply to add to
the usual LDA energy density functional a pairing term
$g(\rho(\bbox{r}))|\nu_c(\bbox{r})|^2$ with a density dependent ``bare
coupling constant $g(\rho(\bbox{r}))$'', extracted from homogeneous
infinite matter calculations. For the descriptions of many systems
(e.g. nuclei, fermionic atomic condensates, $^3$He and neutron matter)
a term linear in $|\nu_c(\bbox{r})|^2$ will most likely suffice.
However, as we already know from the Landau--Ginzburg theory, terms
proportional to $|\nu_c(\bbox{r})|^4$ might become relevant and in
such a case the energy density functional should be generalized
appropriately. Irrespective of the specific functional dependence of
the energy density functional on the anomalous density
$\nu_c(\bbox{r})$, the emerging Kohn--Sham equations will be local and
the ultraviolet divergence in the pairing field will have exactly the
same character as the one studied here and consequently, can be dealt
with using the same approach.

I thank DoE for financial support and G.F. Bertsch and P.--G. Reinhard
for discussions. The very warm hospitality of N. Takigawa in Sendai
and the financial support of JSPS were very helpful while writing the
final version of this work.

%--------------------------------------------------------------------
%--------------------------------------------------------------------

%\newpage

\end{document}